\documentclass[traditabstract, a4paper]{aa}

\usepackage{amsmath}
\usepackage{txfonts}
\usepackage[T1]{fontenc}
\usepackage[latin9]{inputenc}
\usepackage{amstext}
\usepackage{graphicx}
\usepackage{esint}
\usepackage[authoryear]{natbib}
\usepackage{float}
\usepackage{multirow} 
\usepackage{rotating}

\makeatletter
\usepackage{url}
\usepackage{natbib}

\def \cnvphi {\underset{\scriptscriptstyle \phi}{*} }
\def \cnvphit {\stackrel{*}{\scriptscriptstyle \phi} }

\def \cnvsky {\underset{\scriptscriptstyle lm}{*} }
\def \cnvskyt {\stackrel{*}{\scriptscriptstyle lm} }

\def \cnvfull {*}
\def \cnvfullt {*}

\makeatother

\usepackage[english]{babel}
\begin{document}

\title{Faraday synthesis}
\subtitle{The synergy of aperture and rotation measure synthesis} 

\author{M.R.~Bell\inst{\ref{inst:mpa}} and T.A.~En\ss lin\inst{\ref{inst:mpa}}}
\institute{Max Planck Institute for Astrophysics, Karl-Schwarzschild-Str. 1, 85741 Garching, Germany \email{mrbell@mpa-garching.mpg.de}\label{inst:mpa}}

\date{Received \textbf{???} / Accepted \textbf{???}} 

\abstract{We introduce a new technique for imaging the polarized radio sky using
interferometric data. The new approach, which we call Faraday synthesis,
combines aperture and rotation measure synthesis imaging and deconvolution
into a single algorithm. This has several inherent advantages over
the traditional two-step technique, including improved sky plane resolution, 
fidelity, and dynamic range. In addition, the direct visibility-
to Faraday-space imaging approach is a more sound foundation on which
to build more sophisticated deconvolution or inference algorithms.
For testing purposes, we have implemented a basic Faraday synthesis imaging software package
including a three-dimensional CLEAN deconvolution algorithm. We compare
the results of this new technique to those of the traditional approach using mock
data. We find many artifacts in the images made using the traditional
approach that are not present in the Faraday synthesis results. In
all, we achieve a higher spatial resolution, an improvement in dynamic
range of about 20\%, and a more accurate reconstruction of
low signal to noise source fluxes when using the Faraday synthesis technique.}


\maketitle

\section{Introduction}

Rotation measure (RM) synthesis, introduced by \citet{brentjens_faraday_2005},
is a technique that makes use of the Faraday effect to improve the sensitivity of polarimetric
observations by combining data over wide ranges in frequency. RM synthesis allows for the separation of polarized sources along
the line of sight (LOS) by decomposing the observed polarized
emission into parts originating from different Faraday depths (in the simplest case, Faraday depth is equivalent to RM), allowing
one to generate a 3D representation of the polarized sky. While the
Faraday depth axis cannot be mapped to a physical depth, the Faraday
depth information can be of significant scientific value since the
Faraday depth traces the projected strength and orientation of magnetic
fields along the LOS.  A more detailed introduction to RM synthesis is provided later.

The RM synthesis technique has only recently become viable, owing to
the availability of broadband receivers in the next generation of
radio telescopes such as the Expanded Very Large Array (EVLA), the
upgraded Westerbork Synthesis Radio Telescope (WSRT), and the pathfinder projects leading to the Square
Kilometer Array (SKA) such as the Low-Frequency
Array (LOFAR). Recently, there have been many successful applications
of RM synthesis, and interest is rapidly increasing as new radio telescopes
are being commissioned. Applications have included studies of the
diffuse polarized emission in the Perseus field \citep{de_bruyn_diffuse_2005,brentjens_wide_2010}
and the Abell 2255 field \citep{pizzo_deep_2010}, analysis of the
polarized emission in nearby galaxies in the WSRT-SINGS survey \citep{heald_westerbork_2009},
and the detection of a shell of compressed magnetic fields surrounding
a local HI bubble \citep{wolleben_antisymmetry_2010}. The RM synthesis
technique will play a critical role in several upcoming polarization
surveys, e.g. POSSUM \citep{gaensler_possum_2010}, GMIMS \citep{wolleben_gmims_2009},
and future surveys with LOFAR. RM synthesis is very useful for studying magnetism.  For instance, \citet{bell_caustics_2011} have shown that prominent asymmetric features in RM synthesis images known as Faraday caustics, which are related to LOS magnetic field reversals, can be used to study the structural and statistical properties of magnetic fields. 

In addition to the considerable interest in applications,
there has been recent interest in improving RM
synthesis imaging techniques. The RMCLEAN deconvolution algorithm was introduced
by \citet{heald_westerbork_2009} and was quickly adopted,
owing to its simplicity and similarity to techniques used in aperture
synthesis imaging. \citet{frick_wavelet-based_2010} have proposed
a wavelet-based RM synthesis technique. With this approach, one obtains
a decomposition of the size scale of structures in addition to their
Faraday depth location. There has also been growing interest in applying
compressed sensing to RM synthesis \citep{li_compressed-sensing-rmsynth_2011,andrecut_sparse-rmsynth_2011}.
Compressed sensing is a method of reconstructing signals that are
sparse in some set of basis functions. If the signal is sufficiently
sparse, it can be reconstructed using fewer
measurements than indicated by the Nyquist-Shannon sampling theorem. Since
compressed sensing techniques make use of wavelet bases, but are
implemented in a noise-aware fashion, they can be expected to be superior
to pure wavelet based methods.

These new imaging techniques have thus far focused on the problem
of one-dimensional (1D) reconstruction. However,
the product of RM synthesis is a function not only of Faraday depth, but
also of position on the sky, making its reconstruction an inherently
three-dimensional (3D) problem. In many cases, RM synthesis is performed
on sky brightness images that have been produced from radio interferometric
data. Observations performed with a radio interferometer sample the
aperture plane rather than the image plane, and this is done at many
different frequencies. This data space is the 3D Fourier space representation
of the polarized sky brightness as a function of Faraday depth. Imaging algorithms should ideally make
use of the entire data space to inform the reconstruction at each
pixel, since the information about the sky brightness at each pixel
is spread throughout Fourier space. However, the current approach is to perform
the imaging in a piecewise fashion, first reconstructing the 2D sky
plane images at each frequency before doing 1D RM synthesis imaging
along each LOS. Therefore, each step of the traditional imaging approach
is done with a limited subset of the data, which will reduce the overall sensitivity and degrade fidelity
in the final image.

In this paper, we introduce a new technique for imaging the Faraday
spectrum directly from radio interferometric data. We call this technique
\emph{Faraday synthesis}. Using this approach, one images the polarized emission
as a function of sky position and Faraday depth from the
visibility data itself, rather than using the traditional piecewise prescription. Faraday synthesis is a natural extension of the aperture
synthesis plus RM synthesis techniques that provides improvements
in image fidelity and sensitivity. 

We note that a similar technique was briefly discussed in \citet{pen_GMRT-eor_2009},
although it was not considered in any detail, nor was it compared
to the traditional approach to RM synthesis imaging. Furthermore, deconvolution
was not considered. 

With the advent of RM synthesis the concept of rotation measure,
defined to be the amount that the observed polarization angle changes
as a function of frequency, has become somewhat outdated. With RM synthesis, one does not measure RMs, but instead reconstructs the polarized
intensity as a function of Faraday depth. In the simplest case, where
a single, discrete source of polarized emission is positioned behind
a Faraday rotating medium, the RM is equal to the Faraday depth. In
all other cases this is not true. In general, RM cannot be used as a proxy for
Faraday depth, and the full distribution of polarized brightness as a function of Faraday depth is the most appropriate
quantity to study. Therefore, we avoid use of the term RM to describe
this new method, and instead call it Faraday synthesis. Throughout
the remainder of this paper, RM synthesis will refer to
the LOS imaging method developed by \citet{brentjens_faraday_2005}.
The traditional practical approach of first imaging individual frequencies
using 2D aperture synthesis techniques and then reconstructing the
LOS brightness distribution on a pixel-by-pixel basis will be referred to
as 2+1D imaging, in contrast to Faraday synthesis, which we will often
refer to as 3D imaging.

In Sec. \ref{sec:Synthesis-imaging} we briefly review the theories
of aperture and RM synthesis imaging. In Sec. \ref{sec:Faraday-synthesis}
we introduce the Faraday synthesis imaging technique. In Sec. \ref{sec:Proof-of-concept}
we describe the proof of concept software that we have implemented,
and in Sec. \ref{sec:Tests} we compare test results obtained by imaging
mock data using both the 3D and 2+1D techniques. We conclude in Sec.
\ref{sec:Discussions-and-conclusions} with a summary and discussion
of our results.

\section{Synthesis imaging\label{sec:Synthesis-imaging}}

In this section we review the fundamentals of aperture and RM synthesis
imaging before showing how they can be performed simultaneously in
Faraday synthesis imaging. Reviews are included here for completeness
and to highlight the assumptions that are typically made and the limitations
that result.

\subsection{Aperture synthesis\label{sec:aperture_synth}}

We can not possibly provide a complete review of the theory of aperture
synthesis. We only wish to review those aspects that are most relevant
to the current work. For a comprehensive treatment, the reader is
referred to \citet{2001_thompson_interferometry}.

With an interferometer, one measures not the sky brightness directly,
but rather a collection of discrete samplings of the aperture plane.
These samples are complex quantities, typically referred to as visibilities,
denoted as $V$. The visibilities are the correlated voltage output
of pairs of antennas. For a narrow-band observation,
they are related to the sky brightness distribution, $I$, by
\begin{align}
V(u,v,w,\lambda)= & \iintop_{-\infty}^{\infty}\frac{dl\, dm}{\sqrt{1-l^{2}-m^{2}}}\, I(l,m,\lambda) \notag \\
& e^{-2\pi i\left[ul+vm+w\left(\sqrt{1-l^{2}-m^{2}}-1\right)\right]}.\label{eq:vis_sky_relation_full}
\end{align}
The coordinates $(u,v,w)$ are spatial frequency coordinates, or the
distance between pairs of antennas, measured in numbers of wavelengths.
The coordinate $u$ measures the distance in the cardinal North-South
direction, while $v$ measures the distance in the East-West direction.
The coordinate $w$ points in the direction of the phase reference
position on the sky. The $(l,m)$ coordinates are direction cosines
relative to the $(u,v)$ coordinates. The wavelength at which the
visibilities are measured is given by $\lambda$.

This relationship can be simplified to a two-dimensional (2D) Fourier
transformation in two circumstances. The first is in the case of an
East-West oriented array such as the WSRT. In this case, the telescopes move through a plane such that $w=0$ as the Earth
rotates. The
second case is when only a small patch of the sky is being imaged,
such that $w\left(\sqrt{1-l^{2}-m^{2}}-1\right)\approx0$. For the
time being, we assume that we are looking at a small patch of the
sky and henceforth neglect this $w$-term. 

A radio telescope is not equally sensitive to the entire sky. The
sky brightness distribution is attenuated by the antenna power pattern,
$A$, which is commonly referred to as the primary beam. Including
this effect, the visibilities are related to the sky brightness distribution
via the relationship

\begin{equation}
V'(u,v,\lambda)=\iintop_{-\infty}^{\infty}dl\, dm\, A(l,m,\lambda)I(l,m,\lambda)\, e^{-2\pi i\left(ul+vm\right)}.\label{eq:vis_sky_relation_2d}
\end{equation}

In reality, as mentioned above, only discrete locations in the aperture
plane are sampled. The measured visibilities, $\widehat{V}$,
are related to the true visibilities by 
\begin{equation}
\widehat{V}(u,v,\lambda)=S(u,v,\lambda)V'(u,v,\lambda).\label{eq:true_to_measured_visibility}
\end{equation}
The sampling function $S$ can be represented as
\begin{align}
S(u,v,\lambda)= & W(u,v,\lambda)\sum_{i}\delta\left(u-\frac{\mathbf{b_{i}\cdot\hat{x}}}{\lambda}\right) \notag \\
& \delta\left(v-\frac{\mathbf{b_{i}\cdot\hat{y}}}{\lambda}\right)\delta\left(\lambda-\lambda_{i}\right)\label{eq:uv_sampling_fn}
\end{align}
where $\mathbf{b}=b_{x}\hat{x}+b_{y}\hat{y}$ is the distance between
two antennas, known as the baseline length, and the unit vectors $\mathbf{\hat{x}}$
and $\mathbf{\hat{y}}$ point toward the North and East, respectively.
The function $W$ allows for the inclusion of weighting factors, e.g.
by the inverse of the noise. The $i$ subscript is an index over the
list of discrete values of $\mathbf{b}$ and $\lambda$ for which measurements
have been made.

To recover the sky brightness distribution from visibility data, one
must invert Eqs.~\ref{eq:vis_sky_relation_2d} and \ref{eq:true_to_measured_visibility}. Due to the
sampling function and the presence of noise, it is not possible to
solve for $I$ uniquely. The inverse Fourier transform of $\widehat{V}$
does not give the sky brightness distribution alone, but rather
\begin{align}
I_{\textrm{D}}(l,m,\lambda) = & \iintop_{-\infty}^{\infty}du\, dv\,\widehat{V}(u,v,\lambda)\, e^{2\pi i\left(ul+vm\right)}\notag \\
 = & \iintop_{-\infty}^{\infty}du\, dv\, S(u,v,\lambda)V'(u,v,\lambda)\, e^{2\pi i\left(ul+vm\right)}\notag \\
 = & B_{\textrm{sky}}(l,m,\lambda) \cnvsky \left[A(l,m,\lambda)I(l,m,\lambda)\right],\label{eq:2d_dirty_image}
\end{align}
where $\cnvskyt$ denotes convolution in the $l$ and $m$ plane. The image $I_{\textrm{D}}$ is commonly
referred to as the \emph{dirty image}. The \emph{dirty beam,} $B_{\textrm{sky}}$,
is 
\begin{equation}
B_{\textrm{sky}}=\iintop_{-\infty}^{\infty}du\, dv\, S(u,v,\lambda)e^{2\pi i\left(ul+vm\right)}.
\end{equation}

Due to the extended structure of the dirty beam, the sky image obtained
from Eq.~\ref{eq:2d_dirty_image} has a limited dynamic range and
an unphysical appearance. Deconvolution is required
to recover a reasonable approximation of $I$ and to detect
weaker features that are obscured by artifacts associated with the
bright sources. By far, the most commonly used deconvolution algorithm
is the CLEAN algorithm, first introduced by \citet{hogbom74} and
later improved by \citet{clark80} as well as \citet{schwab_1984}.
We describe the CLEAN algorithm later when discussing the 3D implementation
that has been included in our proof-of-concept software.  

In the simplest case, when $A$ is independent of baseline and time, the effect of the primary
beam can be removed by simply dividing by a known beam pattern. In
doing so, the flux scale will be normalized across the sky and the
noise level will increase as a function of distance from the pointing
center. In general, however, $A$ (and other so-called direction dependent effects) 
can depend on both baseline and time, and in this case one must use something like the
A-projection algorithm described by \citet{bhatnagar_aprojection_2008}.

\subsection{RM synthesis}

Faraday rotation is a birefringence effect where the plane of polarization
of a plane-polarized wave is rotated as it passes through a magneto-ionic
medium. The right- and left-circularly polarized components of the
plane wave propagate at different speeds through the medium causing
a relative phase shift, and therefore a rotation of the polarization
plane. The amount of rotation incurred is given by
\begin{equation}
\chi(\lambda^{2})=\chi_{0}+\phi\lambda^{2},\label{eq:PA_rotation}
\end{equation}
where $\chi$ is the observed position angle at wavelength $\lambda$,
and $\chi_{0}$ is the position angle at the source of emission. The
quantity $\phi$, known as the \emph{Faraday depth}, is given by
\begin{equation}
\phi(z)=\left(0.81\textrm{ rad/m}^{2}\right)\intop_{0}^{z}\left(\frac{dz'}{\textrm{pc}}\right)\left[\frac{n_{e}(z')}{\textrm{cm}^{-3}}\right]\left[\frac{B_{z}(z')}{\mu\textrm{G}}\right],
\end{equation}
where $B_{z}$ is the LOS component of the magnetic field, and $z$
is the distance along the LOS. The number density $n_{e}$ includes
both thermal electrons and positrons, which are given as negative
and positive values, respectively. Faraday depth is distinct from
RM, which we define following \citet{burn_depolarization_1966} to
be
\begin{equation}
RM(\lambda^{2})\equiv\frac{\partial\chi(\lambda^{2})}{\partial\lambda^{2}}.
\end{equation}
In the simplest case, when a single point source of polarized emission
sits behind a Faraday rotating medium, the RM measured for the source
is equal to $\phi$. In all other cases, these quantities differ.
In the general case of mixed Faraday rotating and synchrotron emitting
media, the observed polarized intensity originates from a range of
Faraday depths, and the RM varies as a function of $\lambda^{2}$.
The full polarized intensity as a function of Faraday depth, as obtained using RM synthesis,
is required for an opportunity to study the intrinsic properties
of the various sources along the LOS. 

The polarized intensity as a function of sky position and wavelength
is a complex quantity that is usually defined in terms of the Stokes
parameters to be

\begin{equation}
P(l,m,\lambda^{2})=Q(l,m,\lambda^{2})+iU(l,m,\lambda^{2}).
\end{equation}
An important insight from \citet{burn_depolarization_1966} is that this
is related to the polarized intensity as a function of Faraday depth,
$F$, by 
\begin{equation}
P(l,m,\lambda^{2})=\intop_{-\infty}^{\infty}d\phi\, F(l,m,\phi,\lambda^{2})\, e^{2i\phi\lambda^{2}}.\label{eq:P_to_F}
\end{equation}
We refer to $F$ as the Faraday spectrum. The complex phase term reflects that the position angle of the polarized emission originating
at every $\phi$ location is rotated according to Eq.~\ref{eq:PA_rotation}.

Equation \ref{eq:P_to_F} is similar to a Fourier transformation,
and the ability to invert this relationship is desirable, but
two things prevent this. First, it is not possible to
sample $P$ for all values of $\lambda^{2}$. One can only achieve
a limited coverage of $\lambda^{2}>0$, and of course cannot measure
$\lambda^{2}\leq0$ at all. This problem is addressed by the RM synthesis
technique introduced by \citet{brentjens_faraday_2005}, where the
inversion of Eq.~\ref{eq:P_to_F} is treated in much the same way
as the inversion of Eq.~\ref{eq:2d_dirty_image}. Second, the spectral
dependence of $F$ prevents one from inverting Eq.~\ref{eq:P_to_F}.
As addressed by \citet{brentjens_faraday_2005},
the inversion is possible assuming that the $\lambda^{2}$ and
$\phi$ dependent parts of $F$ are separable, which means that the
emission at all $\phi$ values along a LOS is produced with the same
frequency spectrum (up to normalization), i.e. 
\begin{equation}
F(l,m,\phi,\lambda^{2})=f(l,m,\phi)s(l,m,\lambda^{2}).\label{eq:F_factorizable}
\end{equation}
In general, of course, this assumption is not valid. Consider the case where
two sources lie along the line of sight, each having different emission spectra. 
By factoring the Faraday spectrum as above we introduce an error
into the image of $f$. \citet{brentjens_faraday_2005} point out that 
such errors do not effect the Faraday depth of a source, and only have a relatively 
minor effect on the flux density.  In their simulations, using a bandwidth that was 
17\% of the central frequency, an absolute error of 1 in the spectral index 
corresponds to an error of less than 5\% in flux density.  This error should 
be more pronounced with increased bandwidth, since the change in flux density over 
the frequency range becomes greater.

Assuming Eq.~\ref{eq:F_factorizable} applies, the dirty Faraday
spectrum, $f_{D}$, is recovered by 
\begin{align}
f_{\textrm{D}}(l,m,\phi) = & \intop_{-\infty}^{\infty}d\lambda^{2}S_{\lambda^{2}}(l,m,\lambda^{2})\frac{P(l,m,\lambda^{2})}{s(l,m,\lambda^{2})}\, e^{2i\phi\lambda^{2}}\notag \\
= & B_{\phi}(l,m,\phi) \cnvphi f(l,m,\phi) \label{eq:dirty_fs_2+1D}
\end{align}
where $\cnvphit$ is a convolution with respect to $\phi$.
The sampling function, $S_{\lambda^{2}}$, is defined to be
\begin{equation}
S_{\lambda^{2}}(l,m,\lambda^{2})=W_{\lambda^{2}}(l,m,\lambda^{2})\sum_{i}\delta(\lambda^{2}-\lambda_{i}^{2}),
\end{equation}
where $\lambda_{i}^{2}$ are the discrete values of wavelength at
which measurements have been made, and $W_{\lambda^{2}}$ is a weighting
term similar to that in Eq.~\ref{eq:uv_sampling_fn}. The so-called
RM spread function, or the dirty beam in $\phi$-space, $B_{\phi}$,
is given by
\begin{equation}
B_{\phi}(l,m,\phi)=\intop_{-\infty}^{\infty}d\lambda^{2}S_{\lambda^{2}}(l,m,\lambda^{2})\, e^{2i\phi\lambda^{2}}.
\label{eq:dirty_beam_rmsynth}
\end{equation}

Equation~\ref{eq:dirty_fs_2+1D} shows that, like with aperture
synthesis, the product of RM synthesis is a convolution between
the true brightness distribution (the Faraday spectrum in this case) and a dirty beam or point-spread function.
Again, this dirty beam has structure that extends well beyond the
source location. Therefore deconvolution is necessary in order to
recover faint sources in the Faraday spectrum and obtain more physically
realistic results.

\subsection{2+1D imaging of the Faraday spectrum}

RM synthesis is a 1D imaging procedure that is applied to each LOS
independently. With aperture synthesis imaging, one obtains the sky
brightness distribution across the plane of the sky at a single frequency. Applying RM synthesis
to such images requires that the variation of the sky brightness as
a function of frequency must be determined at every location in the
image plane. 

The sampling of the $uv$-plane, $S$, varies as a function of frequency.
The images obtained from Eq.~\ref{eq:2d_dirty_image} at different
wavelengths will be sensitive to different spatial frequencies and
have different resolutions. This complicates measurement of the spectral
properties of the sky brightness because changes due to the sampling
function can be confused with real variation of the sky brightness
distribution. This problem is approximately overcome by applying different
weighting and $uv$-plane tapering schemes to the data such that the
$uv$-coverage is made to be roughly the same at each
frequency. 

The usual practice of applying the RM synthesis technique to interferometric
data involves several steps. We refer to this process as 2+1D imaging:
\begin{itemize}
\item Calibrate the visibility data.
\item Weight and taper the data such that the resolution of the images is
roughly constant with frequency. Some compensation is also needed
if there is significant flux missing due to the usual gap in the center
of the $uv$-plane.
\item Compile a series of deconvolved Stokes Q and U images at each frequency.
Deconvolution is almost always performed using the CLEAN algorithm
and the CLEAN model components are convolved with a so-called restoring
beam, i.e. a Gaussian profile representing the resolution of the image
(determined from the main peak of the dirty beam). The same restoring
beam is used for all frequencies.
\item Stack the images. Pixel-by-pixel, the polarized intensity as a function
of frequency is read from the maps, corrected for spectral variation,
$s$ (determined by measuring the spectral index from total intensity
maps), and Fourier transformed. 
\item Perform further $\phi$-space deconvolution.
\end{itemize}
The result is a 3D cube of data representing $f(l,m,\phi)$. 

There are two immediate problems with this procedure. First, the necessity to downweight data to approximately match resolutions
between images at different frequencies can lead to significant problems.
This cannot be done perfectly and any variation in the polarized intensity
arising from the differing resolutions produces a shift in the Faraday
depth of the emission, thereby introducing systematic errors into
the process. Second, Faraday synthesis is performed on maps that have
already been processed using non-linear, ad-hoc deconvolution algorithms.
Artifacts introduced into the images by said algorithms
will be compounded during RM synthesis. 

We now introduce a new approach that is a natural extension of the
typical synthesis imaging procedures described above, and does not
suffer from the problems of the 2+1D imaging technique.

\section{Faraday synthesis\label{sec:Faraday-synthesis}}

In this section we show that it is possible to directly relate the
Faraday spectrum to the visibilities of the linearly polarized emission.
First, we decompose the Faraday spectrum into parts that relate directly
to the intensity distributions of the two Stokes parameters $Q$ and
$U(l,m,\lambda^{2})$.

Equation \ref{eq:P_to_F} can be rewritten as
\begin{equation}
Q+iU=\intop_{-\infty}^{\infty}d\phi\,\left(F_{Q}+iF_{U}\right)e^{2i\lambda^{2}\phi}
\end{equation}
where we have decomposed $F$ into two complex valued terms, $F_{Q}$
and $F_{U}$, that relate directly to the Stokes $Q$ and $U$ brightness
distributions, respectively. Therefore,
\begin{equation}
Q=\intop_{-\infty}^{\infty}d\phi\, F_{Q}e^{2i\lambda^{2}\phi},
\end{equation}
and the same relation holds for Stokes $U$, as well. We note that
because $Q$ and $U$ are real, $F_{Q}$ and $F_{U}$ are generally
complex and Hermitian in $\phi$, i.e. $F_{Q}^{*}(l,m,\phi)=F_{Q}(l,m,-\phi)$.
We note further that $F_{Q}$ and $F_{U}$ are not the local Stokes
$Q$ and $U$ brightnesses at location $(l,m,\phi)$, but simply auxiliary
variables useful for the formalism. If we assume that $F$ can be
factored as in Eq.~\ref{eq:F_factorizable}, then
\begin{align}
Q(l,m,\lambda^{2})= & s(l,m,\lambda^{2})\intop_{-\infty}^{\infty}d\phi\, f_{Q}(l,m,\phi)e^{2i\lambda^{2}\phi} \notag \\
= & s(l,m,\lambda^{2})q(l,m,\lambda^{2}).
\end{align}

We will now work only with Stokes $Q$, but we note that the following
expressions also hold for Stokes $U$ and $F_U$. Following Eq.~\ref{eq:true_to_measured_visibility},
the measured Stokes $Q$ visibility, $\widehat{V_{Q}}$, is

\begin{align}
\widehat{V_{Q}} = & S\iintop_{-\infty}^{\infty}dl\, dm\, AQ\, e^{-2\pi i\left(ul+vm\right)}\notag \\
 = & S\iintop_{-\infty}^{\infty}dl\, dm\, Asq\, e^{-2\pi i\left(ul+vm\right)}
\end{align}
where $S$ is defined as in Eq.~\ref{eq:uv_sampling_fn}. The relationship
between the Stokes visibilities and the correlator output depends
on the type of antenna feeds that are used. For linearly polarized
feeds 
\begin{align}
V_{Q} = & V_{\textrm{XX}}-V_{\textrm{YY}}\notag \\
V_{U} = & V_{\textrm{XY}}+V_{\textrm{YX}},
\end{align}
where $V_{XX}$, $V_{YY}$, etc. represent the visibilities from cross-correlations
of feeds having X or Y perpendicularly oriented dipoles. For circularly
polarized antenna feeds, 
\begin{align}
V_{Q} = & V_{\textrm{RL}}+V_{\textrm{LR}}\notag \\
V_{U} = & -i(V_{\textrm{RL}}-V_{\textrm{LR}}),
\end{align}
where $V_{RL}$ and $V_{LR}$ represent the visibilities from the
cross-correlation between feeds of right and left, or left and right
circular polarization, respectively. 

We now define $a(l,m,\phi)$ and $\sigma(l,m,\phi)$ to be the representations
of the primary beam and spectral dependence of the Faraday spectrum
in Faraday space, respectively, i.e. 
\begin{equation}
A(l,m,\lambda^{2})=\intop_{-\infty}^{\infty}d\phi\, a(l,m,\phi)e^{2i\lambda^{2}\phi}
\end{equation}
and 

\begin{equation}
s(l,m,\lambda^{2})=\intop_{-\infty}^{\infty}d\phi\,\sigma(l,m,\phi)e^{2i\lambda^{2}\phi}.
\end{equation}
Using the convolution theorem in between $\phi$ and $\lambda^{2}$,
we combine the expressions above to find that the Faraday spectrum
is related to the measured visibilities by
\begin{equation}
\widehat{V_{Q}}=S\iiintop_{-\infty}^{\infty}dl\, dm\, d\phi\, \left( a \cnvphi \sigma \cnvphi f_{Q} \right) e^{-2\pi i\left(ul+vm-\frac{\lambda^{2}}{\pi}\phi\right)},
\end{equation}
where $\cnvphit$ denotes a 1D convolution with respect to $\phi$ along each LOS.
This expression can be inverted to give the dirty image for the Faraday
spectrum
\begin{align}
\left(a \cnvphi \sigma \cnvphi f_{Q}\right)_{\textrm{D}} = & \iiintop_{-\infty}^{\infty}du\, dv\, d\lambda^{2}\,\widehat{V_{Q}}e^{2\pi i\left(ul+vm-\frac{\lambda^{2}}{\pi}\phi\right)}\notag \\
 = & B \cnvfull (a \cnvphi \sigma \cnvphi f_{Q}),\label{eq:3D_inversion}
\end{align}
where
\begin{equation}
B(l,m,\phi)=\iiintop_{-\infty}^{\infty}du\, dv\, d\lambda^{2}\, S(u,v,\lambda^{2})e^{2\pi i\left(ul+vm-\frac{\lambda^{2}}{\pi}\phi\right)},
\end{equation}
and $\cnvfullt$ is a full 3D convolution in $l$, $m$, and $\phi$. We note that the 3D and 1D convolution operations do not commute. 
We can see that by inverting the visibilities, we do not directly
recover the Faraday spectrum, but rather the Faraday spectrum convolved
with $B$ in 3D, and $\sigma$, and $a$ in 1D along each LOS. In order to recover $f_{Q}$, one must first perform a 3D deconvolution using the CLEAN algorithm, for example. After the 3D deconvolution, deconvolution of $a$ and $\sigma$ can be achieved by performing a 1D inverse Fourier transform into $\lambda^2$-space along each LOS, dividing by $A$ and $s$, and then Fourier transforming back into $\phi$-space.  

The beam pattern $A$ is usually known to high precision and
is often represented by an analytic function parameterized in $l$,
$m$, and $\lambda^{2}$. In addition, a map of $s$ will be required. This can be obtained by measuring the spectral variation of total intensity maps along each LOS.  In many circumstances, it may be sufficient to simply assume that $s$ is independent of $l$ and $m$, since the errors introduced by using the wrong form for $s$ are often quite small as previously discussed.

Imaging software that implements the 3D inversion given by Eq.~\ref{eq:3D_inversion}
would avoid the complications described for traditional, 2+1D imaging. We eliminate the need to match
$uv$-coverage at all frequencies because the 3D dirty beam, $B$,
is constructed from the full 3D sampling function. We also avoid
the possibility of compounding errors through the process of deconvolving
images that have already once been deconvolved. As a result, the fidelity
of the images produced using Faraday synthesis should be improved
over those made using the 2+1D technique. In principle, the 3D approach
will result in images that have higher dynamic range than those obtained
using the 2+1D approach because we are able to use all data across
the full bandwidth during imaging and deconvolution.

We have presented a simplified description of the Faraday spectrum
measurement process above. This will already work
quite well in many circumstances, notably for narrow-field observations
without significant direction dependent effects, but in general additional steps will be required. For instance, when the $w$-term in Eq.~\ref{eq:vis_sky_relation_full} can not be ignored, the $w$-projection algorithm of \citet{cornwell_w_2005} has 
been shown to be very effective in reducing imaging errors.  This algorithm makes
use of the fact that the multiplication of the $w$-term and the sky brightness
in the image plane is a convolution in visibility space.  The $w$-dependent visibilities are
projected onto the $w=0$ plane using the convolution kernel, and then a 2D Fourier transform 
can be used to recover the sky brightness distribution.  This algorithm can be applied
to the case of Faraday synthesis without modification. It should also
be possible to apply the A-projection algorithm of \citet{bhatnagar_aprojection_2008}, 
which corrects for direction-dependent beam effects in a manner similar to the $w$-projection algorithm.

\section{Proof of concept implementation\label{sec:Proof-of-concept}}

To compare the 3D approach to polarization imaging with the traditional
2+1D approach, we have implemented a proof of concept 3D imaging and deconvolution
software package called \texttt{fsimager}. For deconvolution, a 3D CLEAN algorithm has been implemented
because it is by far the most common deconvolution method used in
radio astronomical imaging, and with it we can make the most direct
comparison between the two techniques. 

The CLEAN algorithm \citep{hogbom74} is a non-linear, iterative deconvolution
routine. The algorithm makes the implicit assumption that the sky
is composed of point sources distributed throughout a mostly blank
field. Over the last decades, it has been shown to work quite well even
for fields that do not strictly meet this criterion. In brief, the
procedure calls for iteratively building up a model of the sky by
locating the peak of the image, adding a point source to the model
at the location of the peak and with some fraction of its strength,
and subtracting from the image the new model point convolved with
the dirty beam. This is repeated until one can add no further flux to the sky
model, i.e. when one is CLEANing the noise.  A complete description of the 3D CLEAN algorithm that we have implemented is given in Appendix \ref{sec:3DCLEAN}.

\begin{figure}[t]
	\begin{centering}
	\resizebox{\hsize}{!}{\includegraphics{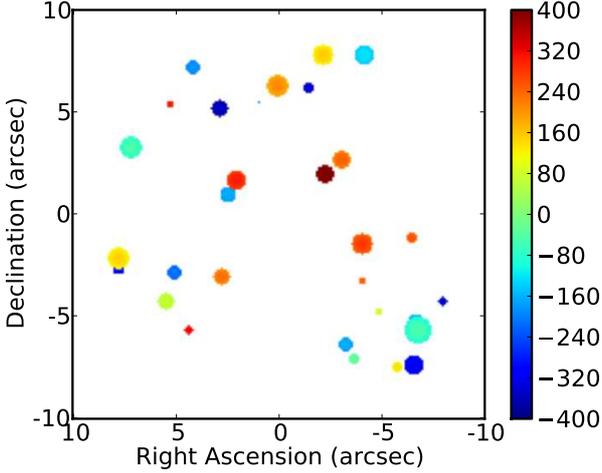}}
	\par\end{centering}
	\caption{The sky-plane distribution of the 30 point sources in the model. 
	The color scale indicates the Faraday depth, given in rad/m$^{2}$, and 
	the area of each circle is proportional to the log of the polarized flux.\label{fig:source-distribution} }
\end{figure}

Radio astronomical imaging relies heavily on Fourier transforms, and
because the number of pixels in the 3D image of the Faraday spectrum
is large, fast Fourier transforms (FFTs) are required for computational
feasibility. Visibility data is not collected on a regularly spaced
grid in $(u,v,\lambda^{2})$-space. To use the FFT algorithm, the
data must be interpolated onto regularly spaced grid points prior
to processing. For this we employ a well known procedure known as
\emph{gridding} that is used extensively in aperture synthesis imaging
as well as in medical imaging. Specifically, we have implemented an
algorithm described by \citet{beatty_gridding_2005}. In brief, we
convolve the data with a Kaiser-Bessel window function (KBWF) and
sample the result on a regular grid in $(u,v,\lambda^{2})$-space.
After this procedure, one is able to perform a FFT with the same result achieved by using a discrete Fourier transformation,
to within an arbitrarily small accuracy. The attenuation of the image
plane caused by the convolution with the KBWF is corrected for by
dividing the dirty image by the Fourier transform of the KBWF. There
are a two parameters in the gridding procedure that affect precision
at the cost of computational time. One parameter describes the extent
of the KBWF in visibility space. The second parameter, the so-called
oversampling ratio, is the factor by which the image plane should
be enlarged in order to mitigate problems that occur at the edges
of the image. With the parameters that we have chosen, for the KBWF
to extend over 6 pixels in each of $u$, $v$, and $\lambda^{2}$,
and an oversampling ratio of 1.5, the results are the same as one
would obtain with a discrete Fourier transformation (DFT) to within one part in 10,000. In fact the
dynamic range is only so limited near the edges of the image where
the effects of the convolution with the KBWF are the most dramatic.
Away from the edges of the image, the dynamic range is much higher.
Overall, the dynamic range can easily be improved by changing the
gridding parameters, but this is done at the expense of increased
processing and memory requirements.

\begin{figure*}
	\begin{centering}
	\includegraphics[width=3in]{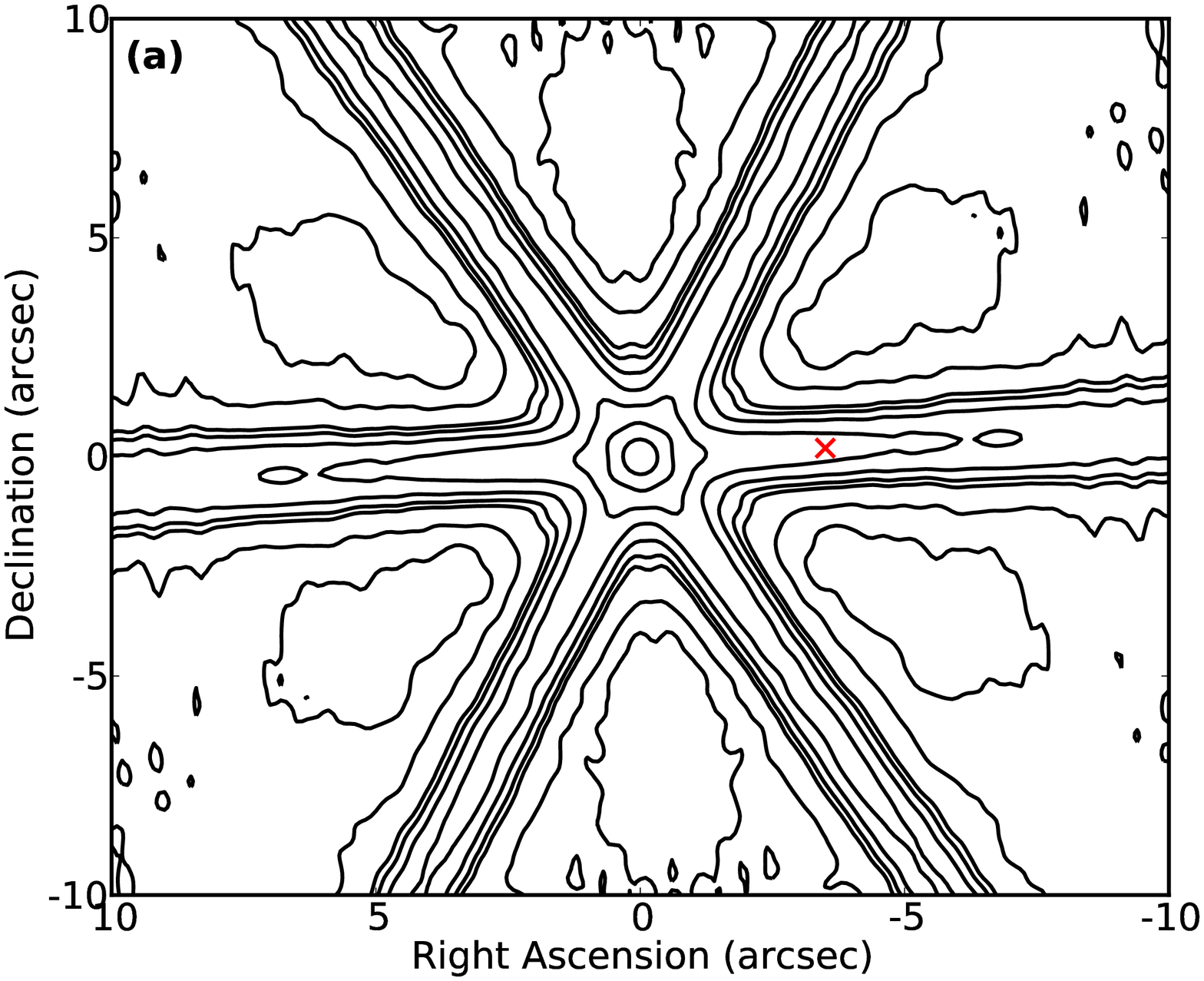}\includegraphics[width=3in]{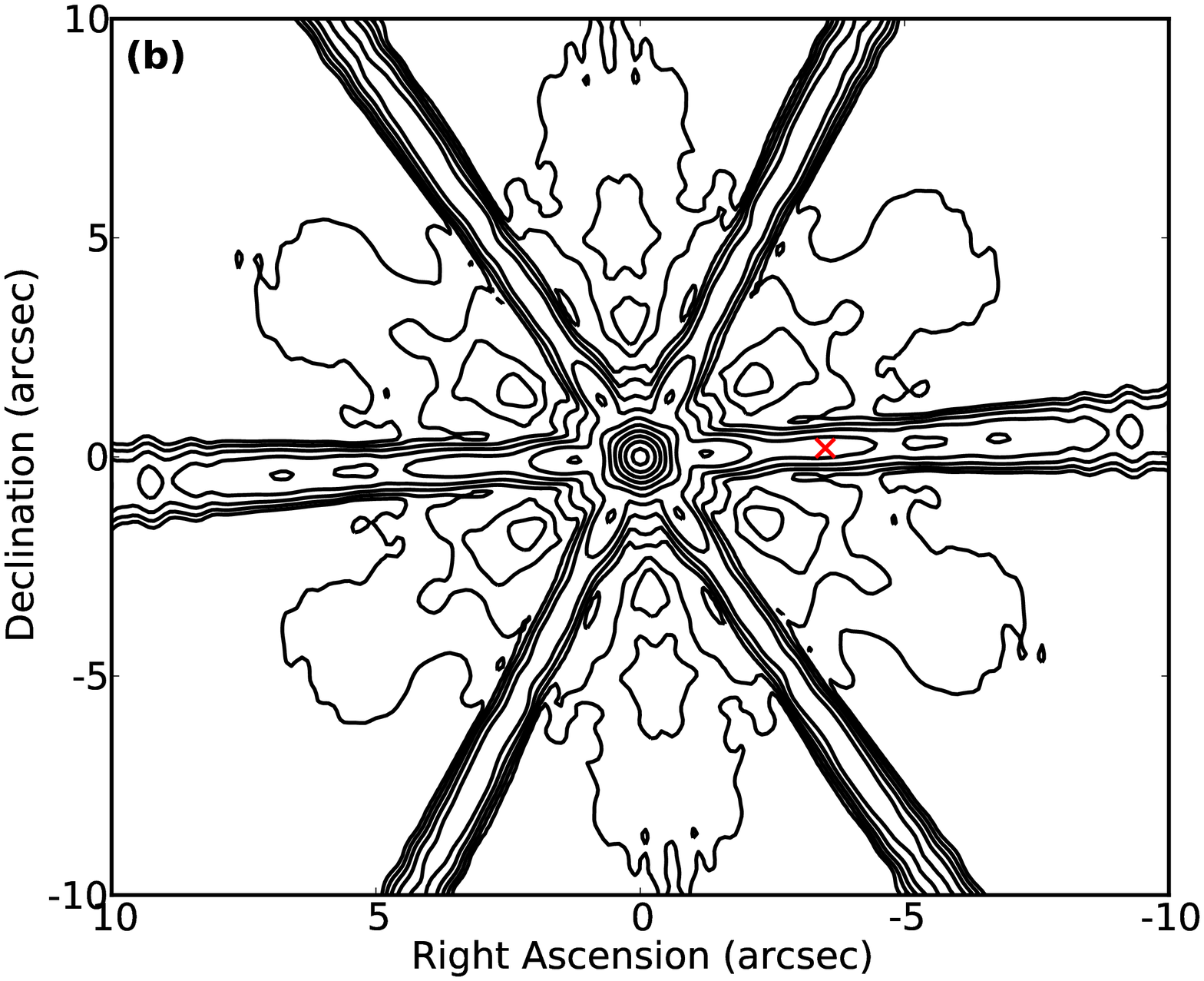}
	\par\end{centering}

	\begin{centering}
	\includegraphics[width=3in]{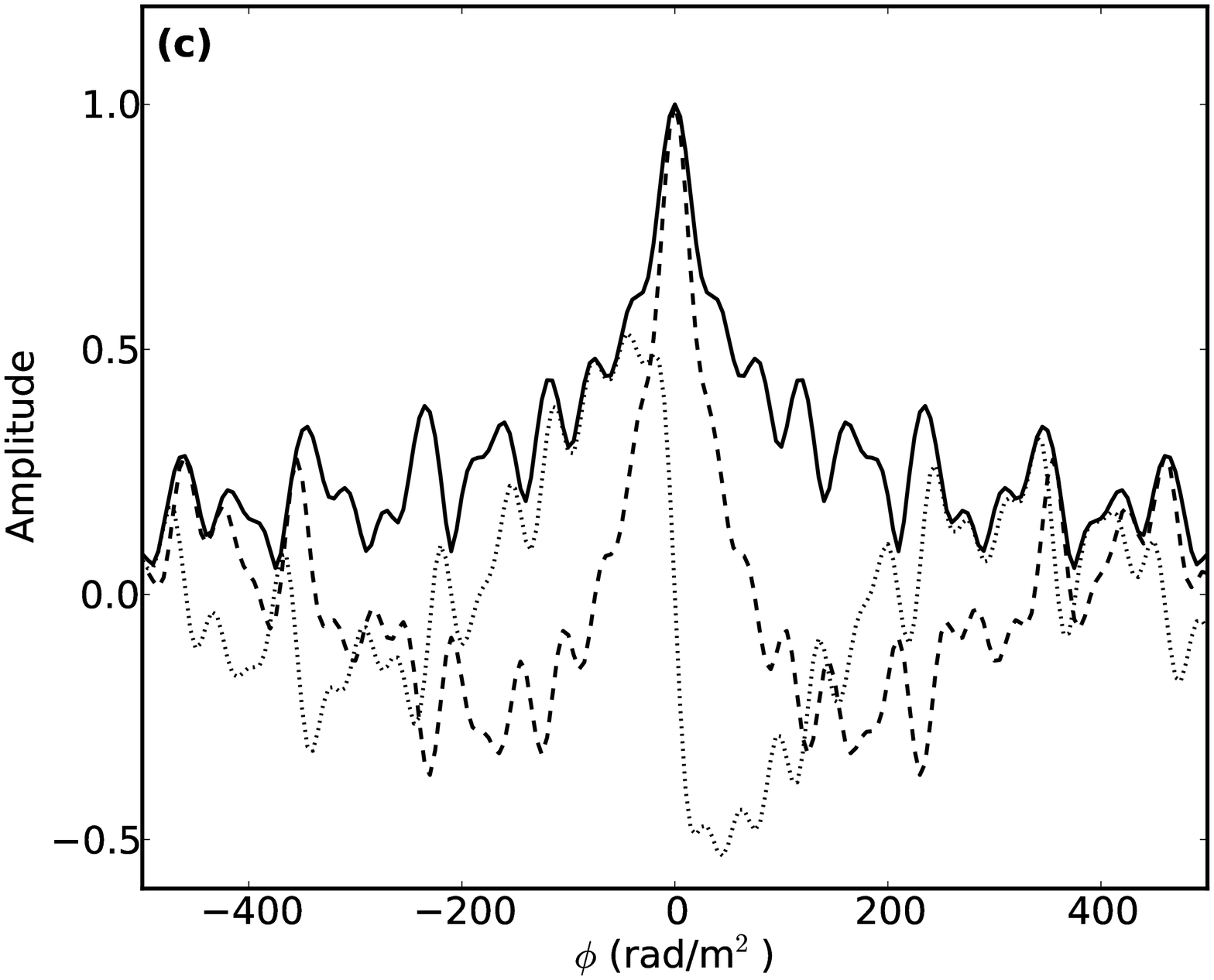}\includegraphics[width=3in]{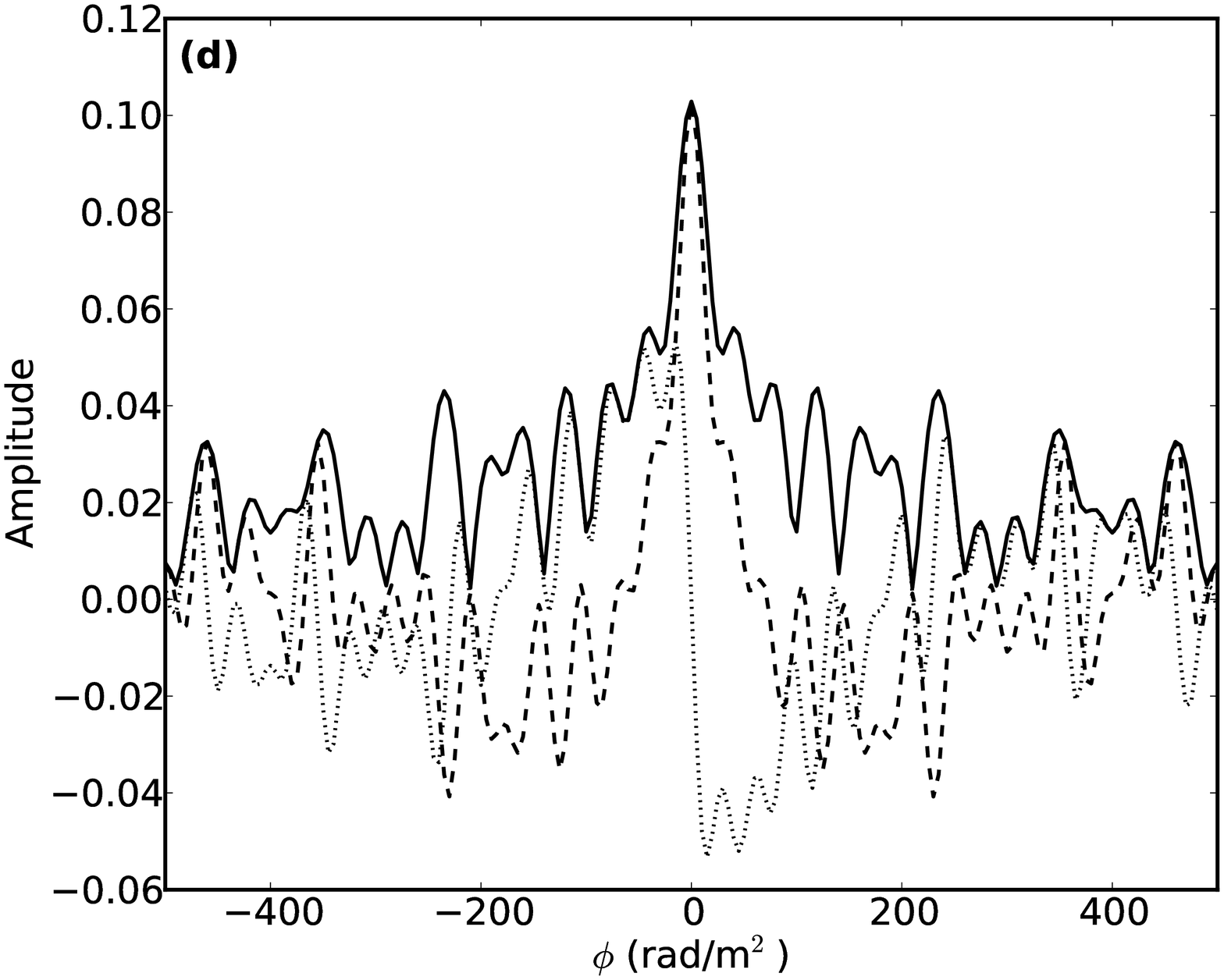}
	\par\end{centering}

	\caption{The 3D dirty beam. a) A 2D slice through the cube in the sky plane
	at $\phi=0$~rad/m$^{2}$. The magnitude of the complex valued beam
	is shown. The bottom contour begins at 0.01 and there is a step of
	2 between levels. Negative contours are marked with dashed lines.
	b) Another view of the beam, now at $\phi=50$~rad/m$^{2}$. c) A
	1D profile through the center of the sky plane at (0'', 0''). d)
	A 1D profile along the LOS at (-3.5'',0.2''), marked with an ``x'' in Figs. a and b. 		The magnitude of the profiles are shown as a solid line, while dashed and dotted lines
	indicate the real and imaginary parts, respectively.\label{fig:3D-dirty-beam}}
\end{figure*}

We have implemented \texttt{fsimager} in Python, thus allowing for
rapid development and testing, but resulting in overall poor performance.
To improve the situation, we have optimized some sub-functions (particularly
the gridding routines) using Cython and in the end the code performs
admirably. We can load a 1GB data set, grid, image, and CLEAN a 16
megapixel image using 500 iterations in about 15 minutes on our 2.4
GHz Core i5 development machine with 8 GB of RAM. The most major limitation
of the current version of the software is that all of the data resides
in memory and this limits the size of the images that can be produced
to the amount of memory that is available on the machine. We have
run all tests on a computer having 64 GB of memory, and are limited
to producing image cubes that are 400 megapixels in size (about 750
pixels per side). This may be sufficient for imaging small fields
of view, but when imaging data from a wide-field, high resolution
instrument (e.g.~LOFAR), this is inadequate.

\begin{figure*}[hp]
	\begin{centering}
	\begin{sideways}\hspace{1.6cm}-205~rad/m$^2$\end{sideways} \includegraphics[width=2.3in, trim=0.5in 0.2in 0.5in 0.49in, clip=true]{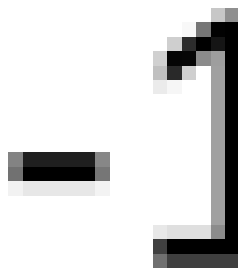}\includegraphics[width=2.3in, trim=0.5in 0.2in 0.5in 0.49in, clip=true]{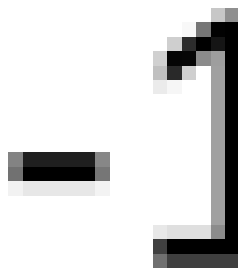}\includegraphics[width=2.3in, trim=0.5in 0.2in 0.5in 0.49in, clip=true]{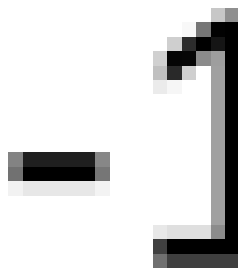}
	\par\end{centering}
	
	\begin{centering}
	\begin{sideways}\hspace{1.6cm}-160~rad/m$^2$\end{sideways} \includegraphics[width=2.3in, trim=0.5in 0.2in 0.5in 0.49in, clip=true]{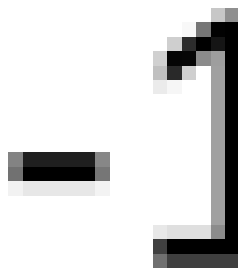}\includegraphics[width=2.3in, trim=0.5in 0.2in 0.5in 0.49in, clip=true]{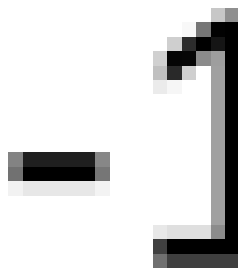}\includegraphics[width=2.3in, trim=0.5in 0.2in 0.5in 0.49in, clip=true]{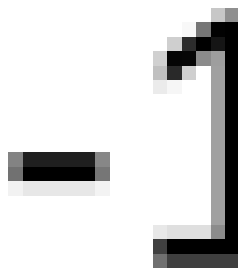}
	\par\end{centering}
	
	\begin{centering}
	\begin{sideways}\hspace{1.7cm}200~rad/m$^2$\end{sideways} \includegraphics[width=2.3in, trim=0.5in 0.2in 0.5in 0.49in, clip=true]{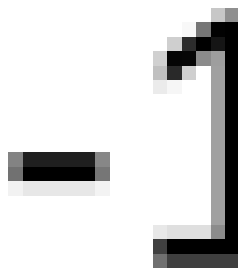}\includegraphics[width=2.3in, trim=0.5in 0.2in 0.5in 0.49in, clip=true]{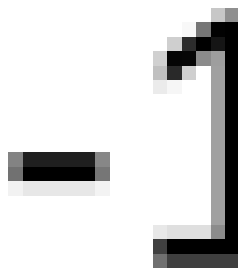}\includegraphics[width=2.3in, trim=0.5in 0.2in 0.5in 0.49in, clip=true]{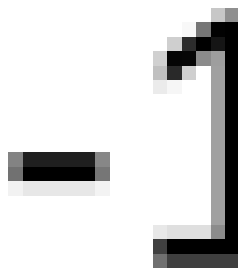}
	\par\end{centering}
	
	\begin{centering}
	\begin{sideways}\hspace{1.7cm}230~rad/m$^2$\end{sideways} \includegraphics[width=2.3in, trim=0.5in 0.2in 0.5in 0.49in, clip=true]{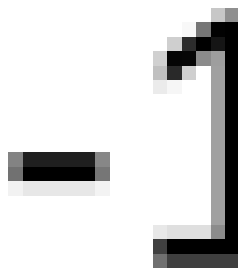}\includegraphics[width=2.3in, trim=0.5in 0.2in 0.5in 0.49in, clip=true]{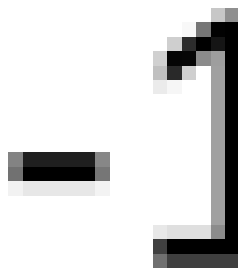}\includegraphics[width=2.3in, trim=0.5in 0.2in 0.5in 0.49in, clip=true]{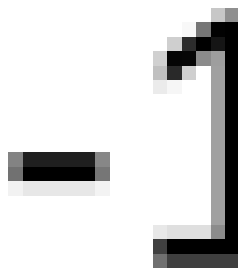}
	\par\end{centering}
	
	\begin{centering}
	\begin{sideways}\hspace{1.7cm}395~rad/m$^2$\end{sideways} \includegraphics[width=2.3in, trim=0.5in 0.2in 0.5in 0.49in, clip=true]{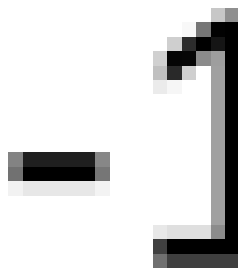}\includegraphics[width=2.3in, trim=0.5in 0.2in 0.5in 0.49in, clip=true]{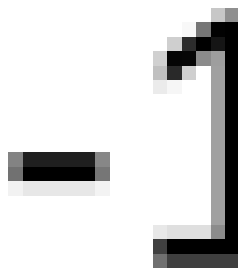}\includegraphics[width=2.3in, trim=0.5in 0.2in 0.5in 0.49in, clip=true]{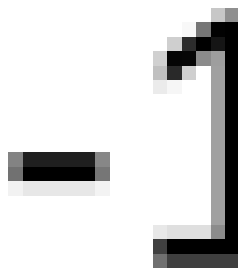}
	\par\end{centering}

	\caption{A comparison between the input model and the two reconstructions at
	several $\phi$ values. In each frame, the greyscale varies
	linearly from $0-50$ mJy/beam, and the contours begin at 50 mJy/beam
	with a factor of 2 between each level. Left: The model image convolved
	with an (0.8''x0.8''x40~rad/m$^{2}$) Gaussian. Middle: The 2+1D
	reconstruction. Right: The \texttt{fsimager} reconstruction. Top to
	bottom:$\phi=-205$, $-160$, $200$, $230$, and 
	$395$~rad/m$^{2}$. \label{fig:reconstructions_-205}}
\end{figure*}

\section{Tests\label{sec:Tests}}

To compare the 3D and 2+1D approaches, we have produced a mock observation
of a set of polarized point sources. In this section, we first describe
the mock data, and then we describe the method that we use to image
the data using the 2+1D method. Finally, we compare the Faraday spectra
produced by each technique.

\subsection{Mock data}

\begin{figure}[]
	\begin{centering}
	\includegraphics[width=2.9in]{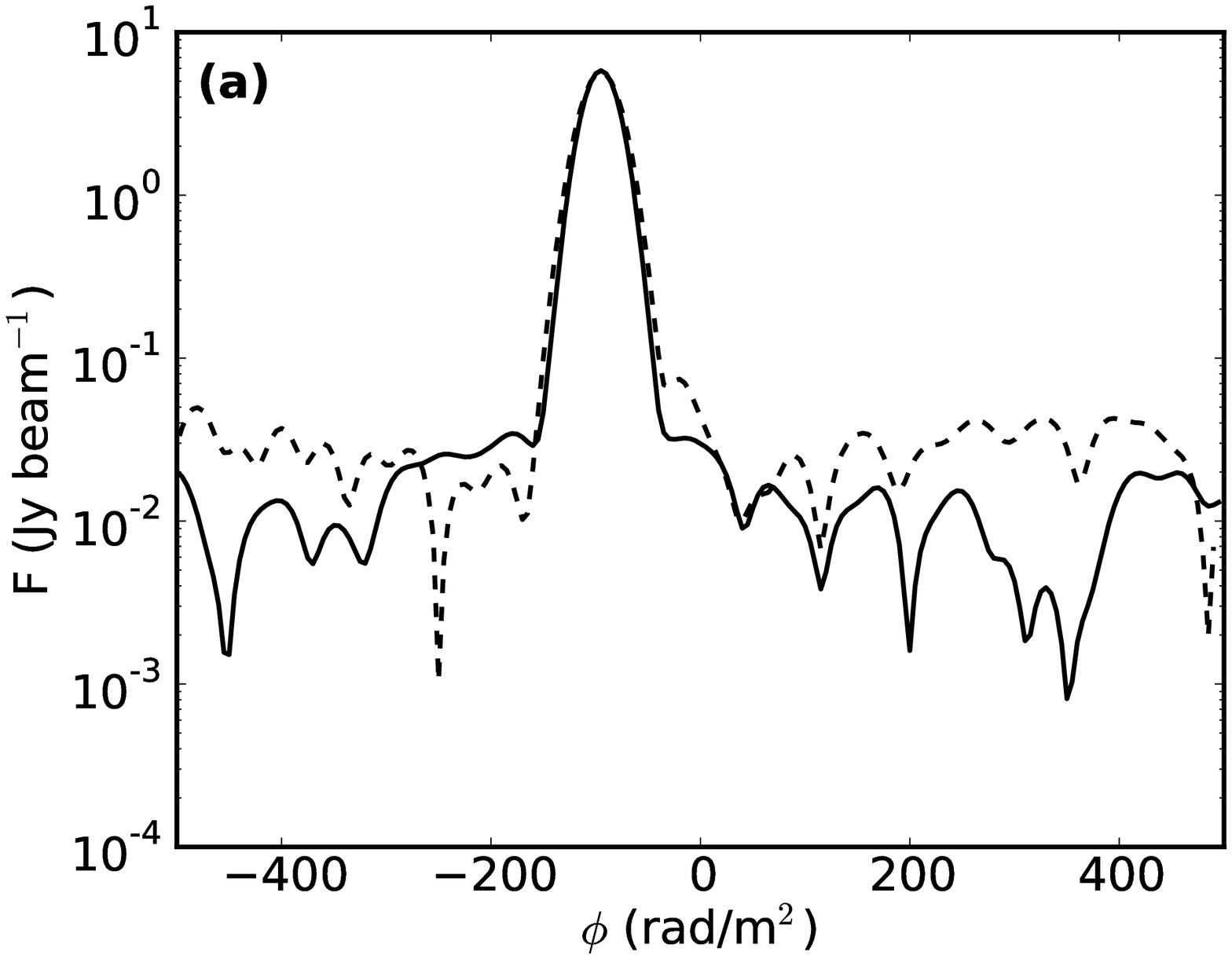}
	\par\end{centering}
	\begin{centering}
	\includegraphics[width=2.9in]{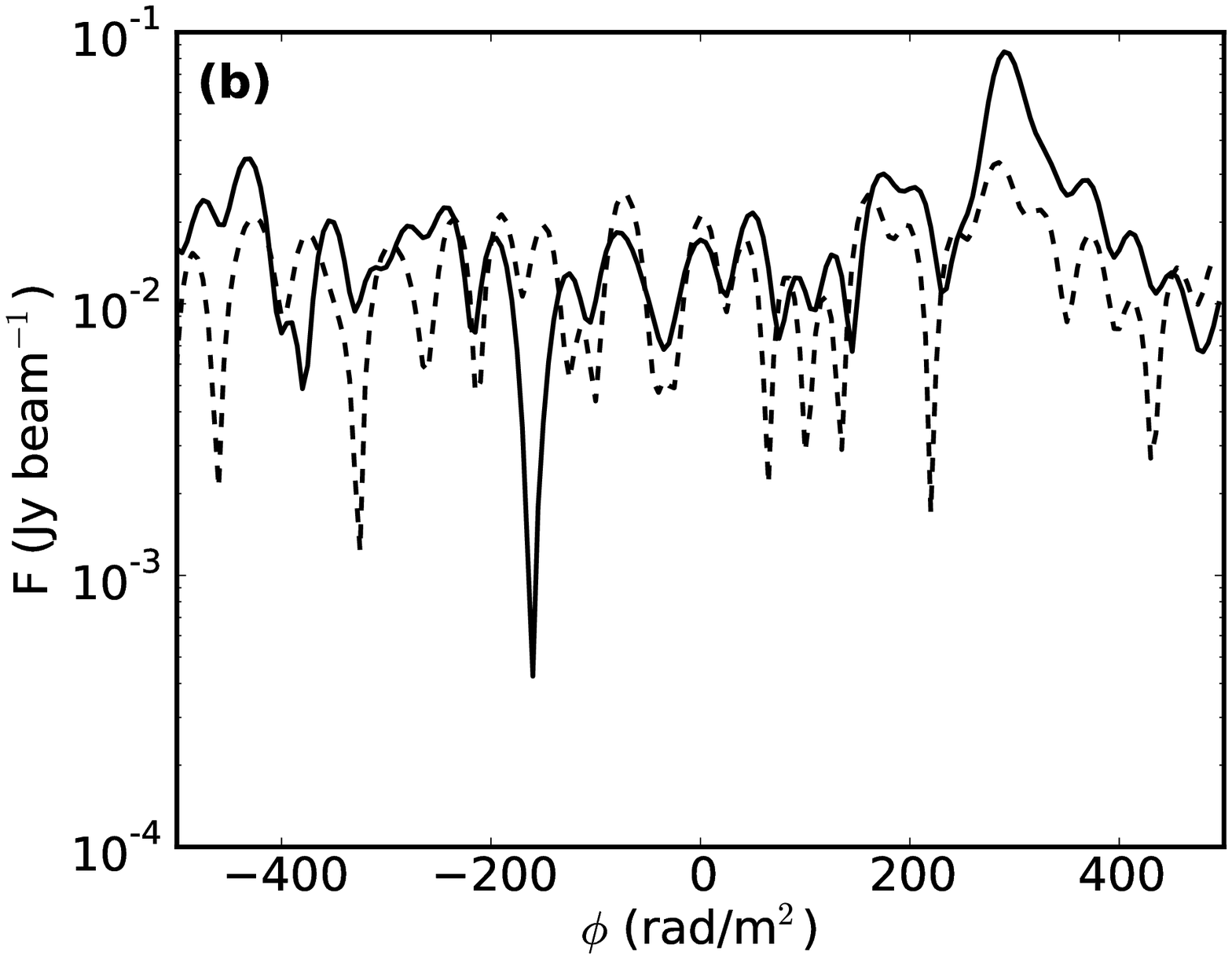}
	\par\end{centering}
	\begin{centering}
	\includegraphics[width=2.9in]{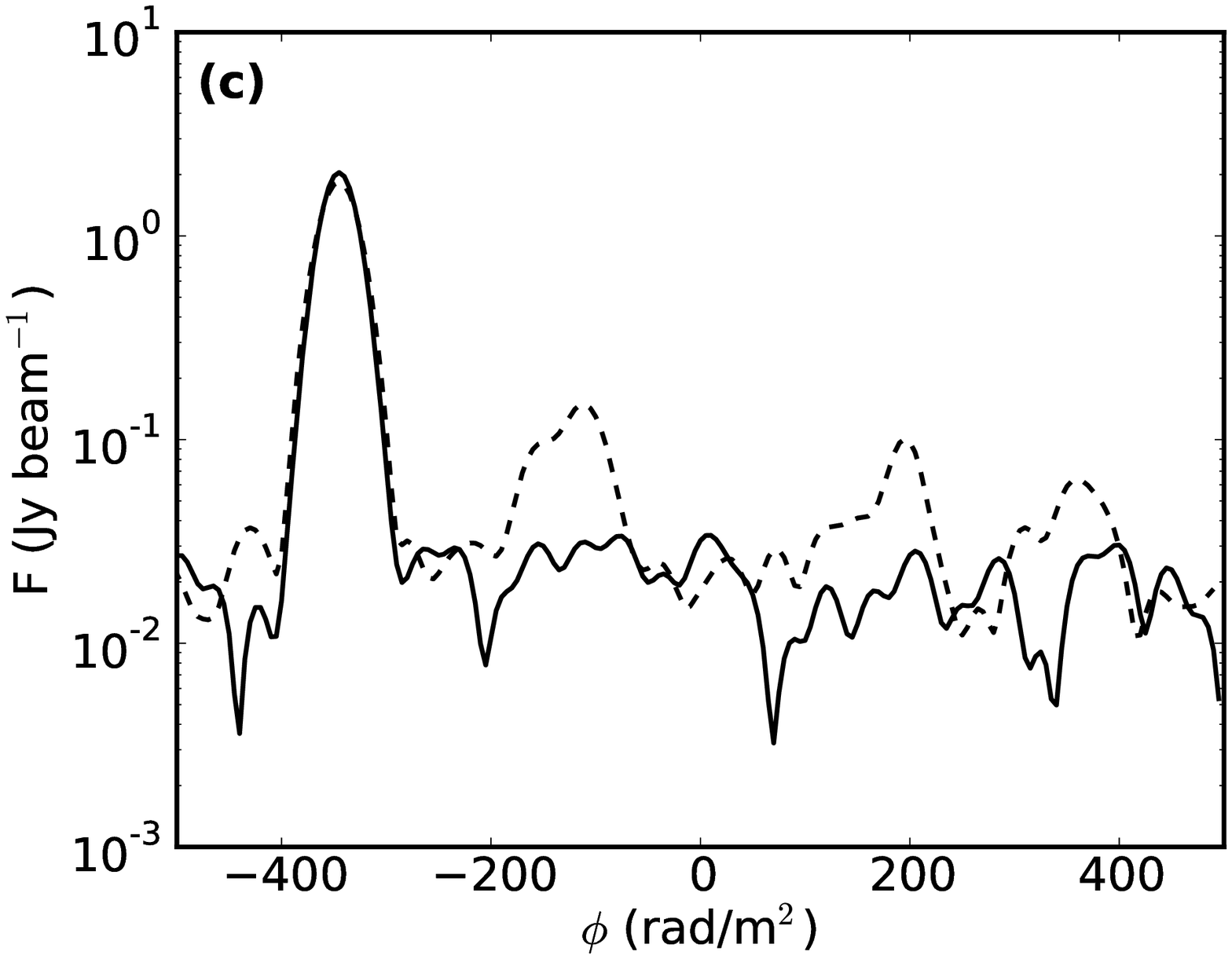}
	\par\end{centering}
	\begin{centering}
	\includegraphics[width=2.9in]{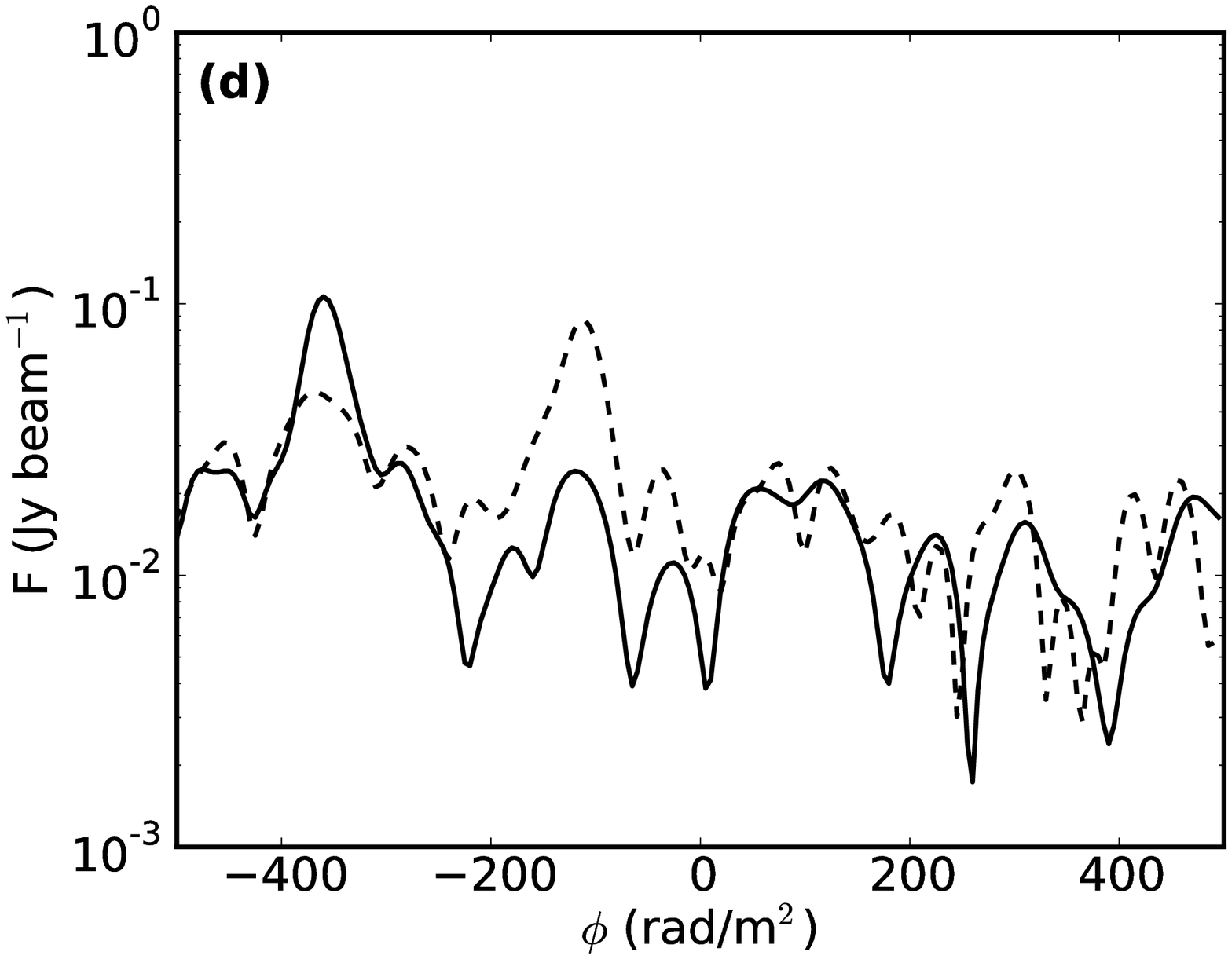}
	\par\end{centering}
	
	\caption{A few LOS profiles through the reconstructed Faraday spectra. Note that the brightness scale is logarithmic. In each,
	the solid and dashed lines indicate the 3D and 2+1D reconstructions,
	respectively. The profiles are taken along the following image plane
	coordinates: a) (7.2'', 3.2''), b) (5.3'', 5.3''), c) (-6.5'',-7.4''),
	d) (-7.9'',-4.3'').\label{fig:A-few-LOS}}
\end{figure}

We create a model Faraday spectrum consisting of a number of
polarized point sources. We choose this simple scenario, though
it may not be the most physically motivated one, because it is what
the CLEAN algorithm was designed for. In these tests we simply want
to study the differences between the two imaging frameworks, not the
merits of the specific imaging algorithm being used. By choosing a
model for which CLEAN is ideally suited, we can concentrate
on the fundamental differences between the 3D and 2+1D techniques
without being concerned about artifacts introduced by using an inappropriate
deconvolution algorithm. 

No spectral variation is included in the model Faraday spectrum, i.e.~$F$
is only a function of $l$, $m$, and $\phi$. Also, we assume that
our field of view is small relative to the size of the primary beam,
and therefore that there is no variation in beam strength throughout
the image plane.

Our model includes 30 polarized point sources, randomly distributed
throughout the volume. The distribution of model sources is shown
in Fig.~\ref{fig:source-distribution}. The color indicates the Faraday
depth, where the scale shown on the right is in rad/m$^{2}$. The
size of the points in the plot indicates the relative linearly polarized
flux of the source. The fluxes range from 0.06 Jy to 64 Jy, thus allowing
for a comparison across a wide range of signal to noise ratios.

We \textquotedbl{}observe\textquotedbl{} the model by taking its Fourier
transformation and then filling the data column of a custom made MeasurementSet
file. The MeasurementSet is created using the \emph{makems} script
that is part of the LOFAR software package. This script creates a
MeasurementSet with user specified time and frequency coverages, and
computes $uv$-coordinates using the antenna table from a pre-existing
MeasurementSet. We use the antenna table from a VLA A-array observation.
The frequency coverage is specified such that our mock data spans
the range from 1-4~GHz, with 64 frequency channels distributed throughout
the range. The total time covered by the observation is 1 hour, with
a 60 second step size between measurements to reduce the file size. 

Gaussian white noise, having a standard deviation of 10~Jy, is added
to real and imaginary parts of the stokes Q and U visibilities separately.
The MeasurementSet file is then either read directly into the 3D software
for gridding, imaging, and deconvolution, or into the CASA software
package for traditional imaging.

\subsection{2+1D imaging procedure}

The 2+1D imaging procedure is conducted by first loading the mock
observation data into CASA where it is then written to UVFITS format
using the task \emph{exportuvfits}. This data is loaded into AIPS
and each channel is independently deconvolved using the IMAGR task. 

In an attempt to match the $uv$-coverage at the different frequencies,
weighting and tapering schemes are employed such that the output image
resolutions are roughly equal. At 1 GHz, the maximum $uv$-spacing
is approximately $125$~k$\lambda$ and the main peak of the synthesized
beam has a FWHM of approximately 1.7''. At 4 GHz, the maximum $uv$-spacing
is 490~k$\lambda$ and the FWHM of the main peak of the synthesized
beam is approximately 0.6''. Without tapering, we will observe large
variations in the sources as a function of frequency simply due to
the dramatic change in resolution.

All but the lowest frequency observations are tapered in the $uv$-plane
with a Gaussian profile having a FWHM of $150$~k$\lambda$. Robust
weighting is used, and the weighting parameter is chosen such that
FWHM of the main peak of the synthesized beam is approximately 1.2''.
After deconvolution in the image plane, all images are restored using
a Gaussian profile that has a FWHM of 1.2''.  We have tried a few other 
weighting and tapering schemes so to achieve even lower resolutions, down 
to a FWHM of 1.7''.  The different schemes made no significant difference 
in the results apart from the resolution of the final image cube.

RM synthesis is performed along each LOS using our own RM synthesis
software that is currently in use on the LOFAR compute cluster for
commissioning. This software implements the RMCLEAN algorithm described
by \citet{heald_westerbork_2009}.

\subsection{Results}

The result of either the 2+1D or 3D imaging procedures is a 3D reconstruction of
of $F(l,m,\phi)$. In Fig.~\ref{fig:reconstructions_-205} we show a side-by-side comparison of
the reconstructed Faraday spectra for a few selected Faraday depths.
In each row, the left panel in each image shows the model image, the middle panel
shows the 2+1D reconstruction and the Faraday synthesis reconstruction
is shown on the right. 

The most obvious difference between the two images is the resolutions. 
In the 3D reconstruction, we use a natural weighting
scheme. The FWHM of the main peak of the 3D dirty beam is roughly
0.8'', 0.8'', and 40~rad/m$^{2}$ in R.A., Dec., and $\phi$, respectively.
Selected image plane slices, and LOS profiles of the Faraday synthesis
derived 3D dirty beam are shown in Fig.~\ref{fig:3D-dirty-beam}. For
the 2+1D imaging, we have in some sense chosen the resolution in the sky plane to be 1.2'' by selecting
a particular weighting scheme, but our choice is limited by the resolution
of the lowest frequency part of the data. We are therefore able to
achieve a higher resolution with the Faraday synthesis technique,
without the need for tapering the long baseline data.

The LOS profile through the center of the 3D dirty beam, shown in Fig.~\ref{fig:3D-dirty-beam}c, is
the same as the usual RM synthesis dirty beam given in Eq.~\ref{eq:dirty_beam_rmsynth}. However, the off-center
profile, shown in Fig.~\ref{fig:3D-dirty-beam}c, is not simply a scaled version of the usual RM synthesis dirty beam.
The structure of the two profiles is quite different. This is because the sidelobes of $B_{\textrm{sky}}$
change as a function of frequency, which leads to structure in $\phi$-space.

We find that the noise level in the 2+1D image is higher than that of the 3D image.  
We measure the noise levels in the two reconstructions by computing the root mean square (RMS)
pixel values in several regions of the image cubes where no sources are located.
In the 3D image cube the noise is 6.06~mJy/beam. The RMS is 20\% higher in the 2+1D image cube, measuring 7.33~mJy/beam.

Both imaging techniques are able to successfully reconstruct the stronger sources in the model image quite well.  The sources above 1 Jy have all been located correctly, and the recovered fluxes are within a few percent of the model fluxes in each case.  In the reconstruction at $\phi=-205$~rad/m$^2$, shown in the first row of Fig.~\ref{fig:reconstructions_-205}, both the 2+1D and 3D reconstructions roughly match the input model. Even the weak sources are detected. This represents one of the better slices of the 2+1D reconstruction, and yet here we can already begin to see problems.  There are some hints of artifacts in the traditionally made image that are not present in the Faraday synthesis result.  The artifacts appear prominently in most other frames of the reconstruction.  

\begin{figure}
	\begin{centering}
	\resizebox{\hsize}{!}{\includegraphics{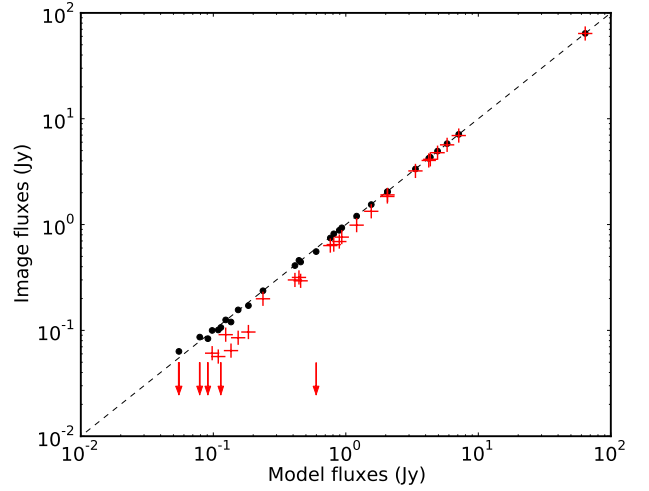}}
	\par\end{centering}
	
	\caption{Reconstructed source fluxes compared to those of the input model.
	Dark circles denote the fsimager reconstruction while light crosses
	denote the 2+1D reconstruction. Arrows indicate upper levels of reconstructed
	fluxes that were not located by our search algorithm.\label{fig:Reconstructed-source-fluxes}}
\end{figure}

Overall, the blank sky in the Faraday synthesis reconstruction is noise-like
throughout the image volume, while in contrast, we find many artifacts
present in the 2+1D reconstruction. The artifacts often appear
as circular outlines of sources at different Faraday depths, like the feature at $\sim$(7'', 3'') in the image at $\phi=-160$~rad/m$^2$. Such features are
present in each of our example image slices above. These circular features
are quite obviously artifacts, but in some cases, for example at $\sim$(-6'',
-7'') in the image at $\phi=200$~rad/m$^2$, false features are
quite strong and not as obviously artificial. One can see a weaker analog of this particular artifact in the 3D reconstruction as well, but most other artifacts in the 2+1D images have no counterpart in the 3D reconstruction.  These erroneous features act
to confuse the detection of weaker sources in the field.

There are artifacts that appear along the LOS profiles of the 2+1D
reconstruction as well. In Fig.~\ref{fig:A-few-LOS}, we show a few
LOS profiles through each image cube. Figure~\ref{fig:A-few-LOS}a shows a profile
through a relatively strong and isolated source at R.A.=7.2'', and
Dec.=3.2'' (c.f. Fig.~\ref{fig:source-distribution}). This source
is well reconstructed by both methods. In the other example profiles,
problems with the 2+1D reconstruction become apparent. The profile shown in Fig.~\ref{fig:A-few-LOS}b is again through a rather isolated but weaker source, at R.A.=5.3'', Dec.=5.3''.  Here we find that the actual source is not prominent compared with the noise in the 2+1D profile, but is easily detected in the 3D reconstruction. 

The other profiles are through more crowded regions of the image cube, where contamination
from nearby sources becomes problematic. The profile in Fig.~\ref{fig:A-few-LOS}c, along R.A.=-6.5''
and Dec.=-7.4'', passes through another relatively strong source.
While the source itself is well reconstructed, three other features
due to nearby sources appear around $\phi=$-150, 200, and 350 rad/m$^{2}$ in the 2+1D reconstruction.  These are not present in the 3D case.
Figure~\ref{fig:A-few-LOS}d shows a profile through a relatively weak source
at R.A.=-7.9'', and Dec.=-4.3''. The source is easily distinguishable
in the 3D profile, but in the 2+1D profile it is completely overshadowed
by the artifact attributed to the source at R.A.=-6.7'', and Dec.=-5.7'',
almost 2'' away. 

These problems occur in part due to the lower resolution
of the 2+1D reconstruction. They are also due to the fact that each
channel in the data is imaged and deconvolved (in 2D) separately,
and therefore with a more limited sensitivity. As a result, residual
artifacts remain in the individual images, and these become apparent
with the increased sensitivity provided by combining data over the
full frequency range. These residuals are then processed during RM
synthesis, and can become quite problematic as we can clearly see
in our examples. 

In each reconstruction, even the weakest model source should be present
at the 10$\sigma$ level in the 3D image and at 8$\sigma$ in the 2+1D case. To locate point sources in each image, we
have a simple routine for locating local maxima that scans through
every pixel and checks whether it is larger than all neighboring pixels.
If so, the location and brightness is recorded. We search through
each image cube in this way to find all local maxima above the 50
mJy/beam level. Ideally we should expect to find 30 points corresponding
to the model sources. In the 3D reconstruction, we find 32 such locations.
All input model sources are located along with two false detections.
In the 2+1D case we locate 147 sources, with 5 input model sources
not detected. A few of the missing sources are not detected because
they are simply not resolved from a nearby stronger source. The others
may be present below the 50 mJy/beam cutoff, but at this point they
are completely indistinguishable from the artificial sources.

The artificial sources in the 2+1D reconstruction are not only much more numerous, but also brighter.  The brightest 
artifact in the 2+1D image cube is 0.113~Jy/beam, compared to 0.062~Jy/beam in the 3D image cube.

Figure \ref{fig:Reconstructed-source-fluxes} compares the fluxes
of the sources found in the two reconstructions to those of the input
model. These have been corrected for the Ricean bias effect described
in \citet{wardle74}. The 3D reconstruction agrees remarkably well
with the model fluxes across the full range of source strengths. This
sky model is ideally suited for the CLEAN algorithm, so this is as
should be expected. The stronger sources are also recovered nicely
in the 2+1D reconstruction, but the strengths of weaker sources are
systematically lower than the input model fluxes.

\section{Discussion and conclusions\label{sec:Discussions-and-conclusions}}

We have described a new approach to imaging linearly polarized visibility
data that we call Faraday synthesis. With this approach, one directly
reconstructs the Faraday spectrum, or the polarized intensity as a
function of Faraday depth, from the visibility data. This takes the
place of the usual approach of first performing aperture synthesis
imaging on the visibility data at each frequency, then performing
RM synthesis along each line of sight independently. 

These two approaches would be equivalent if deconvolution
were not required. With Faraday synthesis, the deconvolution is done
in one step using the entirety of the broad-band data. In contrast, the traditional
approach requires deconvolving the images individually at each frequency.
The sensitivity in the narrow-band images is significantly limited, and residual artifacts
remain in these images that limit the dynamic range and image fidelity achieved during
RM synthesis. Moreover, another deconvolution procedure is
necessary to account for the point spread function due to the incomplete
sampling of the wavelength space. Artifacts introduced by the first
deconvolution algorithm will be compounded during this procedure, further
reducing image fidelity.

Indeed, we found in our proof-of-concept testing that artifacts were
significantly higher in the traditional imaging method than with
Faraday synthesis. The noise was roughly 20\% lower when using the Faraday synthesis technique, and the strongest artifact was about half as bright.

We found that one is able to achieve a better resolution in the final
image using the new approach. The main lobe of the 3D dirty beam was
30\% smaller in the sky plane than that of the traditional method.
With the 2+1D method, stacking of the images at each frequency requires
tapering the visibility data such that the higher frequency images
have the same resolution as those at the lower frequencies. We find that the inaccuracies inherent in the stacking process are a significant source of artifacts in the traditional RM synthesis technique.  Furthermore, this procedure requires severely down-weighting a large
portion of the available data, again limiting sensitivity. With the Faraday synthesis approach, no such down-weighting
is required.

The Faraday synthesis approach of working directly between visibility
and Faraday space is a much better foundation on which to build new
image reconstruction algorithms because with it one is able to accurately describe the instrument response. The effects of the intermediate, non-linear deconvolution procedure can not be easily understood or modeled. Many signal inference techniques, like those derived using Information Field Theory \citep{ensslin_ift_2009,ensslin_gibbs_2010}, depend on a complete description of the instrument response and would need to be built on the framework of Faraday synthesis.

While the CLEAN algorithm has worked quite well in radio astronomy for decades, the implicit
assumption of a sky sparsely populated by point sources is not well suited for the kinds of diffuse polarized signals that one finds, for example, in the polarized emission from the Milky Way. Novel image reconstruction algorithms built using more appropriate constriants or assumptions are likely justified. Such algorithms could make use of statistical correlations inferred from the data, similar to the extended critical filter algorithm developed by \citet{oppermann_extcritfilt_2011}.

\begin{acknowledgements}

This research was performed in the framework of the DFG Forschergruppe
1254 Magnetisation of Interstellar and Intergalactic Media: The Prospects
of Low-Frequency Radio Observations. We thank Henrik Junklewitz, Niels
Oppermann, Marco Selig, Maximilian Uhlig, George Heald, and Ger de
Bruyn for many helpful discussions. We thank Ger van Diepen at ASTRON
for the \emph{makems} software that we use to produce our mock data.

\end{acknowledgements}

\bibliographystyle{aa}
\bibliography{rmsynth_refs}

\appendix

\section{3D RMCLEAN}\label{sec:3DCLEAN}

In our proof of concept software we have implemented
a variant of the CLEAN routine that was introduced by \citet{clark80}.
In this variant the model is populated by searching for peaks in image
space, but the model subtraction is performed in $uv$-space. 

We first produce the dirty images of $F_{Q\textrm{D}}$ and $F_{U\textrm{D}}$.
These are combined into a single complex image according to
\begin{align}
\Re(F_{\textrm{D}}) = & \Re(F_{Q\textrm{D}})-\Im(F_{U\textrm{D}})\notag \\
\Im(F_{\textrm{D}}) = & \Re(F_{U\textrm{D}})+\Im(F_{Q\textrm{D}}),
\end{align}
where $\Re$ and $\Im$ are operators that select the real and imaginary
parts of the image, respectively. We proceed with the CLEAN procedure
using the complex valued $F_{\textrm{D}}$, and the associated visibilities.

The algorithm is performed in two parts, the so-called major and minor
cycles. In the minor cycle, new model sources are located in image
space. The subtraction of the model from the visibility data is performed
during the major cycle. What remains after the model is subtracted
from the visibilities are called the residual visibilities, $V_{\textrm{R}}$. 

Before starting the procedure, we extract a patch from the dirty beam,
$B_{\textrm{patch}}$, for use during the minor cycle. We record the
value $\beta=\max(|B-B_{\textrm{patch}}|)$, where $|\cdot|$ indicates
that we take the magnitude of the complex valued map.

The major cycle includes:
\begin{enumerate}
\item Invert $V_{\textrm{R}}$ to create a residual image of the complex
Faraday spectrum, $F_{\textrm{R}}$. Find $F_{\textrm{lim}}=max(|F_{\textrm{R}}|)$.
\item Start the minor cycle using $F_{\textrm{R}}$ and populate a minor
cycle sky model, $M_{\textrm{minor}}$. The minor cycle is described
below.
\item Upon completion of the minor cycle, inverse Fourier transform the
minor cycle sky model into visibility space, $V_{\textrm{M}}$. 
\item Subtract the model from the residual visibility data, $V_{\textrm{R}}=V_{\textrm{R}}-SV_{\textrm{M}}$.
\item Add $M_{\textrm{minor}}$ to the complete sky model, $M_{\textrm{major}}$.
\item Repeat steps 1-6 until some user defined number of iterations has
been done, or $F_{\textrm{lim}}$ is below a user defined cutoff.
\item Invert $V_{\textrm{R}}$ to produce a final residual image. Add to
this $M_{\textrm{major}}$ convolved with a Gaussian restoring beam.
\end{enumerate}
The minor cycle proceeds as follows:
\begin{enumerate}
\item Find $(l_{\textrm{m}},m_{\textrm{m}},\phi_{\textrm{m}})=argmax(|F_{\textrm{R}}|)$. 
\item Add a point source to $V_{M}$ at $(l_{\textrm{m}},m_{\textrm{m}},\phi_{\textrm{m}})$
with a flux $F_{\textrm{M}}=gF_{\textrm{R}}(l_{\textrm{m}},m_{\textrm{m}},\phi_{\textrm{m}})$,
where $g$ is a user defined gain factor (between zero and one).
\item Subtract $F_{\textrm{M}}B_{\textrm{patch}}$, centered on $(l_{\textrm{m}},m_{\textrm{m}},\phi_{\textrm{m}})$,
from $F_{\textrm{R}}$.
\item Continue as long as $|F_{\textrm{M}}|>\mbox{\ensuremath{\beta}}F_{\textrm{lim}}(1+\frac{1}{N})$
where $N$ is the total number of minor cycle iterations that have
been completed.
\end{enumerate}
We adopt the minor cycle stop condition suggested in \citet{clark80}.
This condition reflects the fact that during the minor cycle we only
subtract a small patch of the dirty beam from the image and therefore
any effects outside of this patch will not be removed from the image.
During the minor cycle, we CLEAN only as deeply as the maximum contribution
of the brightest source to the residual image that is \emph{not} removed
by subtracting the beam patch. Actually, the $(1+\frac{1}{N})$ term
makes this condition even more strict early on in the procedure.

\end{document}